\documentclass[10pt]{article}
\usepackage{amssymb}
\usepackage{graphicx}
\usepackage{epstopdf}

\newcommand{\be}{\begin{eqnarray}}
\newcommand{\ee}{\end{eqnarray}}

\begin{document}

\begin{titlepage}

\title{Implications of primordial black holes on the first stars and 
the origin of the super--massive black holes} 

\author{Cosimo~Bambi$^{\rm 1}$, Douglas~Spolyar$^{\rm 2}$,
Alexander~D.~Dolgov$^{\rm 3,4,5}$, \\ Katherine~Freese$^{\rm 6}$, 
and Marta~Volonteri$^{\rm 7}$}

\maketitle

\begin{center}
$^{\rm 1}$IPMU, The University of Tokyo, 
          Kashiwa, Chiba 277-8568, Japan \\
$^{\rm 2}$Department of Physics, University of California, 
          Santa Cruz, CA 95060, USA \\
$^{\rm 3}$Istituto Nazionale di Fisica Nucleare, Sezione di Ferrara, 
          I-44100 Ferrara, Italy \\
$^{\rm 4}$Dipartimento di Fisica, Universit\`a degli Studi di Ferrara,  
          I-44100 Ferrara, Italy \\
$^{\rm 5}$Institute of Theoretical and Experimental Physics,
          113259 Moscow, Russia \\
$^{\rm 6}$MCTP, University of Michigan, 
          Ann Arbor, MI 48109, USA \\
$^{\rm 7}$Department of Astronomy, University of Michigan, 
          Ann Arbor, MI 48109, USA
\end{center}

\vspace{0.5cm}

\begin{abstract}
If the cosmological dark matter has a component made of small 
primordial black holes, they may have a significant impact on 
the physics of the first stars and on the subsequent formation 
of massive black holes. Primordial black holes would be adiabatically 
contracted into these stars and then would sink to the stellar 
center by dynamical friction, creating a larger black hole which 
may quickly swallow the whole star. If these primordial black 
holes are heavier than $\sim 10^{22}\;{\rm g}$, the first stars 
would likely live only for a very short time and would not 
contribute much to the reionization of the universe. They would instead 
become $10 - 10^3\; M_\odot$ black holes which (depending on subsequent 
accretion) could serve as seeds for the  super--massive black holes 
seen at high redshifts as well as those inside galaxies today.
\end{abstract}

\end{titlepage}

\section{Introduction}

The first stars in the Universe mark the end of the cosmic dark ages,
reionize the Universe, and provide the enriched gas required for later
stellar generations.  They may also be important as precursors to
black holes (BHs) that coalesce and power bright early quasars. The 
first stars are thought to form inside Dark Matter (DM) halos of mass 
$ 10^5 \, M_\odot$--$ 10^6 \, M_\odot$ at redshifts $z=10-50$ 
({Abel} {et~al.} (2002); {Bromm} {et~al.} (2002); 
{Yoshida} {et~al.} (2003)). These halos consist of 85\% DM and 15\% 
baryons in the form of metal-free gas made of H and He.  Theoretical 
calculations indicate that the baryonic matter cools and collapses via 
H$_2$ cooling ({Peebles} \& {Dicke} (1968); {Matsuda} {et~al.} (1971); 
{Hollenbach} \& {McKee} (1979); {Tegmark} {et~al.} (1997)) into a 
single small protostar ({Omukai} \& {Nishi} (1998)) at the center 
of the halo (for reviews see {Ripamonti} \& {Abel} (2005), 
{Barkana} \& {Loeb} (2001) and {Bromm} \& {Larson} (2004)).

It is interesting to study the effects on the evolution of the first
stars due to the large reservoir of DM within which these stars form.  
The first protostars and stars are particularly good sites for this 
investigation because they form inside the highest density environment 
(compared to today's stars): they form at high redshifts (density scales 
as $(1+z)^3$) and in the high density centers of DM haloes. Previously, 
two of us ({Spolyar} {et~al.}(2008)) 
studied the effects of Weakly Interacting Massive 
Particles on the first stars: we found that the annihilation of these 
particles could provide a heat source for the star that stops the 
collapse of the protostar as well as potentially dominates over any 
fusion luminosity for a long time.

In this paper, we consider instead the effects on these first stars of
a different candidate for the DM: Primordial Black Holes (PBHs). 
These are small black holes that may be formed in the very early universe (see the next section
for more detail) 
and may exist in sufficient abundance to provide
the DM seen in the Universe today.  The masses of PBHs that explain the entirety
of the DM range from    ($10^{17}-10^{26}$) g; heavier PBHs up to $1 \, M_\odot$ may provide
still an interesting fraction of the DM.

 We discuss the implications that 
PBH DM would have on the physics of the first stars, the so called 
Population III stars. These stars could range from $\sim$ (1 - few 100) $M_\odot$.  
First, we compare various possible heat sources due to PBHs with the ordinary heat from
stellar fusion of the stars.  For the properties of the Pop III stars, we use results computed by Heger \& 
Woosley.  Specifically, for a 100~$M_\odot$  (10~$M_\odot$) Pop. III star, we take the 
central temperature to be $1.2 \times 10^8$~K ($9.6 \times 10^7$~K), the 
central density  31~g/cm$^3$ ($226$~g/cm$^3$), the radius 7~$R_\odot$ 
(1.2~$R_\odot$), and the stellar fusion luminosity to be
\be 
\label{eq:lstar}
L_* &=& 6.5 \times 10^{39} \, {\rm erg/s} \,\,\,\, (100 \, M_\odot) \, , \\
L_* &=& 4.2 \times 10^{37} \, {\rm erg/s} \,\,\,\, (10 \, M_\odot)  \, .
\label{eq:lstar10}
\ee
We find that the ordinary stellar fusion luminosity dominates over the
 heat sources due to PBHs, which include accretion onto 
the BHs, Hawking radiation, and the Schwinger mechanism. 

Instead, we find the interesting result that PBHs inside the first stars may 
sink to the center and form a single BH, which may accrete very rapidly 
and swallow the whole star.  The phenomenon is relevant for PBHs
heavier than about $10^{22}$~g, because the corresponding timescale 
for dynamical friction turns out to be shorter than the typical
stellar lifetime, while it is less interesting or completely 
negligible for lighter BHs. So, for $M_{PBH} \gtrsim 10^{22}$~g, the lifetimes of 
Pop. III stars may be  shortened, with implications for
reionization of the Universe as well as for the first supernovae.
In addition, since the stars are 
inside much larger haloes, they can in principle accrete even more matter 
(depending on the accretion mechanism).  Thus,  the end--products of the scenario are 
BHs of masses $10-10^5 \, M_\odot$.   These may be the seeds which produced 
the super--massive BHs seen at high redshifts; the Intermediate Mass Black Holes; as well as 
the black holes at the center of every normal 
galaxy today and whose origin is as yet uncertain. Possible mechanisms of 
production of superheavy BHs are reviewed in {Dokuchaev} {et~al.} (2007). 
In addition, although the PBH swallowing the star shortens the star's 
lifetime and its contribution to reionization, the newly formed hole can 
become a new, alternative source of ionizing photons.

The rest of the paper is organized as follows. In Section II,
we briefly review the physics of PBHs, that is, how they can be
formed in the early Universe and what current constraints on their
cosmological abundance are.
In Section III, we discuss the behavior of individual PBHs: how many of
them are expected inside a single star (via adiabatic contraction),
what is the luminosity due to accretion onto the PBHs, and what is the
timescale for their size to double.  We also investigate alternative
mechanisms for generating luminosity by these small PBHs.  Then in
Section IV and V we turn to the most important part of the paper. We 
study the dynamical friction that pulls all the BHs into a single
larger BH at the center of the star, and then watch this single large
BH accrete the entire star surrounding it on a fairly rapid timescale.
We conclude with a discussion in Section VI. 
Throughout the paper, we use units with $c = 1$.

\section{Physics of Primordial Black Holes}

\subsection{Production mechanisms}

PBHs may be formed in the early universe by many 
processes ({Zeldovich} \& {Novikov} (1966), {Hawking} (1971), 
{Carr} \& {Hawking} (1974), {Crawford} \& {Schramm} (1982), 
{Hawking} (1989), {Polnarev} \& {Zembowicz} (1991), 
{Dolgov} \& {Silk} (1993), {Jedamzik} (1997), {Rubin} {et~al.} (2000), 
{Dolgov} {et~al.} (2008)). For a general review, see e.g. {Carr} (2003).  
The earliest mechanism for BH production can be fluctuations in the 
space-time metric at the Planck epoch.  Large number of PBHs can also 
be produced by nonlinear density fluctuations due to inhomogeneous 
baryogenesis at small scales ({Dolgov} \& {Silk} (1993), 
{Dolgov} {et~al.} (2008)). If within some region of space density 
fluctuations are large, so that the gravitational force overcomes the 
pressure, we can expect the whole region to collapse and form a BH. 
In the early Universe, generically, BHs of the horizon size are formed, 
although it is also possible to form much smaller BHs ({Polnarev} \& 
{Zembowicz} (1991), {Hawking} (1989)). BHs can also be produced in first 
and second order phase transitions in the early Universe 
({Crawford} \& {Schramm} (1982), {Jedamzik} (1997)).  Gravitational 
collapse of cosmic string loops ({Polnarev} \& {Zembowicz} (1991), 
{Hawking} (1989)) and closed domain walls ({Rubin} {et~al.} (2000)) 
can also yield BHs.  The masses of PBHs formed in the above mentioned 
processes range roughly from $M_{Pl}$ (BHs formed at the Planck epoch) 
to $M_\odot$ (BHs formed at the QCD phase transition).

The basic picture is that energy perturbations of order one stopped
expanding and recollapsed as soon as they crossed into the
horizon ({Zeldovich} \& {Novikov} (1966), {Hawking} (1971), 
{Carr} \& {Hawking} (1974)).  The maximal mass of PBHs is set by 
the total mass within the cosmological horizon, i.e. 
$M_{hor} = M_{pl}^3/\Lambda^2$ at any given energy scale $\Lambda$ at
which the BH forms.  This is also the expected mass scale of a BH in 
most early Universe scenarios for the production of PBHs (it can be at 
most a factor of $10^{-4}$ smaller ({Hawke} \& {Stewart} (2002))).  Thus 
\be\label{h-mass} M_{PBH} \approx \frac{t_f}{G_N} \approx 5 \cdot
10^{26} \, g_*^{-1/2} \, \left(\frac{\rm 1 \; TeV}{T_f}\right)^2 \;
{\rm g} \, , \ee 
where we assumed a radiation dominated Universe, with $g_*$ the 
effective number of relativistic degrees of freedom and $T_f$ the 
temperature of the Universe at time $t_f$.

\subsection{Observational constraints}

PBHs in the mass range $M_{PBH} \sim 10^{17} - 10^{26}$~g can be good
DM candidates. A number of  constraints restrict the mass to this range.
PBHs with an initial mass smaller than about $5 \cdot 10^{14}$~g are 
expected to be already evaporated due to Hawking radiation; moreover 
their presence in the early Universe can be constrained by observations 
for $M_{BH} \gtrsim 10^9$~g (lifetime $\tau \gtrsim 1$~s) 
({Novikov} {et~al.} (1979)). For $M_{PBH} \sim 10^{15}$~g, there are 
strong bounds as well, at the level of $\Omega_{PBH} \lesssim 10^{-8}$, 
from the observed intensity of the diffuse gamma ray 
background ({Page} \& {Hawking} (1976)), so they may be at most 
a tiny fraction of the non--relativistic matter in the Universe.  For 
larger masses, constraints can be deduced from microlensing 
techniques ({Alcock} {et~al.} (2000); {Tisserand} {et~al.} (2007)) 
and dynamical arguments ({Carr} \& {Sakellariadou} (1999)), which 
exclude the possibility that the whole cosmological DM is made of BHs
heavier than $10^{26}$~g, even if they still may be an important
component. For example, the PBH to DM mass ratio in the Galactic 
Halo would be smaller than 0.04 for PBHs in the mass range
$10^{30} - 10^{32}$~g and than 0.1 for the mass range 
$10^{27} - 10^{33}$~g ({Tisserand} {et~al.} (2007)).

On the other hand, for the mass range $10^{17} - 10^{26}$~g, there are
currently no clear observational methods of detection.  For $M_{PBH}
\sim 10^{17} - 10^{20}$~g, the presence of PBHs can be inferred from
the femptolensing of gamma ray burts ({Gould} (1992), {Nemiroff} \& 
{Gould} (1995), {Marani} {et~al.} (1999)), but 
the constraint is weak, roughly $\Omega_{PBH} \lesssim 0.2$; in addition
it holds only for uniformly distributed DM and is not easy to 
extend to the more realistic case of clumped DM.  The same mass range 
might be covered by future gravitational wave space antennas, from the 
gravitational interaction of PBHs with test masses of the laser 
interferometer ({Seto} \& {Cooray} (2004)), but the expected detection 
rate for LISA is too low and only a further generation of space 
detectors might put non--trivial constraints. According to recent 
work {Abramowicz} {et~al.} (2008), the PBH mass range 
$10^{15} - 10^{26}$~g remains unexplored and thus allowed.
However, further constraints raise the lower bound to roughly
$10^{16} - 10^{17}$~g ({Bambi} {et~al.} (2008a)).

We present results for PBHs with mass $M_{BH} = 10^{24}$~g 
but show the scaling for other PBH masses in the $10^{17}-10^{26}$ 
range.  Our results are qualitatively the same for PBHs of any mass in the 
allowed range.  For heavier PBHs up to e.g. $1 \, M_\odot$, the results
will be somewhat different and discussed in the discussion section.

\section{Primordial Black Holes inside the Star}

In this section we study the behavior of the PBHs inside the star. We
estimate the total mass in these objects, as well as the luminosity
and timescale for accretion onto individual PBHs.

\subsection{Total Mass in PBHs inside the star}
The first stars form at the centers of $10^6 M_\odot$ DM haloes.  As a
starting point we assume an initial Navarro--Frenk--White profile 
({Navarro et al.} (1997)) for
both DM (85\% of the mass) and baryons (15\% of the mass). As the gas
collapses to form a star, it gravitationally pulls the DM (in this
case PBHs) with it.  We use adiabatic contraction ({Sellwood} \& 
{McGaugh} (2005)) to find the resultant dark matter profile inside 
the star ({Spolyar} {et~al.} (2008))
\be
\label{eq:profile}
\rho_{DM} \approx 5 \, (n_b \,\, {\rm cm}^3)^{0.8} \; {\rm GeV/cm}^3 \, ,
\ee 
which is  independent of the nature of DM\footnote{This
is the result of a calculation for DM density in the first stars that
we performed with WIMP dark matter in mind, but exactly the same
result holds for any type of DM including BHs which are orders of magnitude larger.}. 
Here, $n_b$ is the mean baryon density inside the star.
It should be noted that adiabatic contraction is not a relaxation process.  Instead as the 
baryons collapse to form a star, they gravitationally pull the DM will them, so that the DM density
inside the star increases. 
  Hence, the DM evolves on the timescale of the baryons. Ideally, instead of using the 
  adiabatic approximation, it would be desirable to run an N-body simulation. At present this is 
  technically not possible. Regardless, adiabatic contraction should give a reasonable approximation 
  and is widely used formalism\footnote{Our original work on adiabatic contraction in the first stars
  was performed using a very simple assumption of circular orbits only. However, in follow-up work,
 two of us participated in a paper ({Freese et al.} (2009) ) in which we performed an exact calculation including radial 
 orbits.  The results changed by less than a factor of two, so that we feel comfortable using Eq.(\ref{eq:profile}).
 In that same paper we also considered a core alternative to an NFW profile as our starting point for the adiabatic
 contraction and, again, obtained essentially the same result.  Our results for DM densities in the first stars
 appear to be quite robust.}.  In addition,  the formal requirements to apply the adiabatic 
  approximation hold during  most of the evolution of the baryons.  
For a mean baryon number density
$n_b \approx 10^{24}$~cm$^{-3}$, the DM to baryon matter mass ratio of
a typical Pop. III.1 star is at the level of $10^{-4}$.  The number of
PBHs inside the star is roughly
\be
N_{BH} \sim
10^7 \left({10^{24}~g \over M_{BH}}\right ) \left({M_* \over 100~M_\odot}
\right) ,
\ee
where $M_*$ is the mass of the Pop. III star.  More precisely (modeling
the star as an $n=3$ polytrope), we find for a 100~$M_\odot$ 
(10~$M_\odot$) star that the total mass in PBHs is 
\be
\label{eq:mtot}
M_{tot,PBH} &=& 6.3 \times 10^{30} \; {\rm g} \,\,\,\, (100 \, M_\odot) \, , \\
M_{tot,PBH} &=& 4.1 \times 10^{29} \; {\rm g} \,\,\,\, (10 \, M_\odot) \, .
\label{eq:mtot10}
\ee

\subsection{Accretion onto the PBHs from stellar material}

In this paper we study the effects of PBHs on the stars on the main
sequence, once they have fusion proceeding in their cores. The PBH
effects are much more important during this stage than during the
protostellar collapse phase. Since they are
surrounded by a high density stellar environment, they accrete and
emit radiation. As a maximum possible value, the accretion luminosity
for a single PBH 
cannot exceed the Eddington limit 
\be\label{edd-lim} L_E = \frac{4 \pi
G_N M_{BH} m_p}{\sigma_{Th}} = 6.5 \cdot 10^{28}
\left(\frac{M_{BH}}{10^{24} \; {\rm g}}\right) \; {\rm erg/s} \, , \ee 
where
$\sigma_{Th}$ is the Thomson cross section. $L_E$ is the luminosity at
which the outwards radiation pressure compensates the gravitational
attraction and stops the accretion process. 
Clearly, the Eddington luminosity is proportionate to mass. In this case, the mass has been conservatively 
set to the mass of the BH ($M_{BH}$).  In fact, the mass should include the optically thick gas surrounding the BH.
Under this restriction, the maximum stellar
luminosity from PBHs inside one star is realized when 
the accretion luminosity of every BH is at the Eddington limit, i.e. 
\begin{equation}
L_{E,tot} = N_{BH} L_E
\sim 10^{36} \left({M_* \over 100 \; M_\odot}\right) \, {\rm erg/s} .
\end{equation}
Since $L_{E,tot} \propto M_{PBH,tot}$, the upper
bound on the power emitted by PBHs is independent of the BH mass.
This accretion powered luminosity is at least a few orders of
magnitude smaller than the expected stellar luminosity for Pop. III
stars, $4 \cdot 10^{37}$~erg/s ($6 \cdot 10^{39}$~erg/s) for 10 and
100~$M_\odot$ ({Freese} {et~al.} (2008)) stars respectively.  The extra 
heat produced by accretion onto the PBHs inside the star has 
thus a negligible impact on the physics of the star.

As the PBHs accrete more matter and become more massive, the Eddington
limit increases and the BH accretion luminosity becomes more and more
important in the energy balance of the star.  The Bondi accretion rate
is ({Bondi} (1952))
\begin{equation}
\label{bondi-eq} \dot{M}_B = 4 \pi R_B^2 \, \rho_b
\, v = 1.4 \cdot 10^{12} \, \left(\frac{M_{BH}}{10^{24} \; {\rm
g}}\right)^2 \left(\frac{1 \; {\rm keV}}{T}\right)^{3/2}
\left(\frac{\rho_b}{1 \; {\rm g/cm^3}}\right) \, {\rm g/s} \, .  
\end{equation}
Here $R_B = 2 G_N M_{BH} / v^2$ is the Bondi radius.   The quantity $v$ is 
the typical velocity of the particles of the accreting gas with
respect to the BH, and  should account for both the thermal
velocity of the particles $v_p = \sqrt{3T/m_p}$, where $T$ is the 
local temperature of the star, as well as the BH orbital velocity
$v_{BH} = \sqrt{G_N M_*(r)/r}$, where $M_*(r)$ is the stellar mass
within a distance $r$ from the center. Close to the center $v \approx v_p$,
but for large $r$ this relation is no longer true; instead, the BH orbital velocity may
reduce the accretion rate, even by an order of magnitude.  We
here take $v = v_p$ and use the Bondi formula to find an upper 
limit on the accretion rate, recognizing that this value may well 
overestimate the true accretion rate \footnote{Moreover, the 
Bondi formula holds in the ideal case of
perfect spherical symmetry. In realistic situations, the non--zero
angular momentum of the accreting gas and the presence of other
effects (magnetic fields, turbulences, etc.) may diminish the
accretion rate, since $L_a$ must be smaller than $L_E$. The case of
accretion onto BHs is however a complex phenomenon, because BHs have
an event horizon and in principle may be capable of swallowing an
arbitrary amount of matter without exceeding the Eddington
luminosity ({Begelman} (1978), {Begelman} {et~al.} (2008)). We will  
take the Bondi accretion as an upper limit ({Begelman} (1978)).}. 
The differential equation $\dot{M}_{BH} = \alpha M_{BH}^2$ 
has solution \be M_{BH}(t) = \frac{M_0}{1 - \alpha
M_0 t} \, , \ee where $M_0$ is the BH mass at $t = 0$ and 
\be \alpha
M_0 = 1.6 \cdot 10^{-13} \left(\frac{M_{BH}}{10^{24} \; {\rm
g}}\right) \left(\frac{1 \; {\rm keV}}{T}\right)^{3/2} 
\left(\frac{\rho_b}{1 \; {\rm g/cm^3}}\right) \; {\rm s^{-1}} \, 
\ee 
is the inverse of the characteristic accretion time of the BH.  
The accretion time scale is thus not shorter than 
\be\label{timescale} \tau_a \sim 10^5
\left(\frac{10^{24} \; {\rm g}}{M_{BH}}\right) 
\left(\frac{T}{1 \; {\rm keV}}\right)^{3/2} 
\left(\frac{1 \; {\rm g/cm^3}}{\rho_b}\right) \; {\rm yr} \, .  
\ee

It is possible for even a single PBH with $M_{BH} > 10^{24}$~g inside 
the star to eat the entire star. Such a case was discussed in {Begelman} (1978)
in the context of a super--massive star capturing a BH in a bound
orbit.  The current scenario differs due to the fact that we are interested in
the role of PBHs on Pop. III stars and their effects on cosmology 
(e.g. reionization); here the PBHs are thought to comprise at
least some measurable fraction of the DM in the universe and are therefore
present in the haloes containing the Pop. III stars before these even form.
If the PBHs do not comprise the entire DM, then the PBH mass could be larger
than we have discussed heretofore, though contributing only a small fraction
of the critical density.

As we will show below, the maximal accretion
rate computed here is somewhat slower than the rate for 
the formation of a larger BH at the center of the star; all the
effects combined thus lead to a big BH at the center.

\subsection{Other mechanisms for energy release by PBHs}

One may be also concerned about  two other mechanisms in which PBHs can 
release energy: Hawking radiation ({Hawking} (1975,1976)) and positron 
emission ({Bambi} {et~al.} (2008b)).  

\subsubsection{Hawking radiation}
The luminosity due to Hawking radiation is maximal for the smallest 
mass BHs. We thus consider the 
(unrealistic) possibility that all the cosmological DM is made of PBHs 
with mass $M_{BH} = 10^{14}$~g. The Hawking luminosity per BH from 
$\gamma$, $e^\pm$ and $\mu^\pm$ emission is 
$7 \cdot 10^{18}$~erg/s ({Page} (1976)) and their total contribution to 
the power of a 10~$M_\odot$ star would be at the level of $10^{35}$~erg/s, 
roughly 2 or 3 orders of magnitude smaller than the ordinary stellar 
luminosity, $4 \cdot 10^{37}$~erg/s. If the mass of the star were 
100~$M_\odot$, the relative contribution would be smaller, because 
the stellar luminosity increases by a factor 100, while the BH luminosity increases
by a factor 10. Higher Hawking luminosity would demand smaller PBHs.  However, 
if the PBHs had an initial mass $M_{BH}=10^{13}$~g, their lifetime would 
be $\tau < 10^5$~yr, that is much shorter than the age of the Universe 
when first stars formed.  Thus fusion luminosity always dominates over 
the Hawking radiation.

\subsubsection{Schwinger effect}
The second mechanism, positron production due to Schwinger effect at
the BH horizon, has been recently discussed in {Bambi} {et~al.} (2008b).
Because protons are much more massive than electrons, it is much
easier for BH to capture protons. Whereas the protons fall right into
the BH, the electrons interact frequently via Compton scattering on
their way into the BH and are prevented from falling in as
easily. Hence the BH builds up a positive electric charge.  For a BH
mass $M_{BH} < 10^{20}$~g, the electric field at the BH horizon can
exceed the critical value for vacuum stability, i.e. $E_c = m_e^2/e$,
so that electron--positron pairs can be efficiently produced (Schwinger
effect).  Then, electrons are back--captured while positrons escape.
The net result is to convert protons of the surrounding plasma into
150~MeV positrons.  
The accretion rate of protons is ({Bambi} {et~al.} (2008b))
\begin{equation}
\dot N_p = 10^{30} \,
\left({M_{BH} \over 10^{20} 
{\rm g}} \right)^2 \left ( { 1 {\rm keV} \over T} \right )^{3/2} 
\left(\frac{\rho_b}{1 \; {\rm g/cm^3}}\right) \;
{\rm s}^{-1} \, .
\end{equation}
We note that mechanism is not the same as Bondi accretion, 
and that the expression above
is not obtained from eq.~(\ref{bondi-eq}). By contrast, Bondi accretion is the 
accretion of gas where particles collide with one another other, losing
their tangential velocity but gaining radial velocity towards 
the star. This hydrodynamic approximation is applicable if the 
characteristic length scale is larger than the mean free path of particles.
Here, the size of the BH is smaller than the proton mean free path,
$\lambda_p$, and we consider protons at distances $r < \lambda_p$ 
with small velocities, so they are gravitationally bound to the BH.
These protons lose energy by bremsstrahlung or synhrotron 
radiation near the BH and in this sense they are not non-interacting.
The picture is very much diffent from the hydrodynamical one and
the calculations of the proton accretion rate can be found in 
{Bambi} {et~al.} (2008b). Once equilibrium is reached between 
the accretion rate and the Schwinger discharge rate, 
the luminosity per BH is roughly ({Bambi} {et~al.} (2008b)) 
\be
L_{e^+} \sim 3 \times 10^{26} \left(\frac{M_{BH}}{10^{20} \; {\rm g}}\right)^2
\,  \left(\frac{1 \;
{\rm keV}}{T}\right)^{3/2} 
\left(\frac{\rho_b}{1 \; {\rm g/cm^3}}\right)
\; {\rm erg/s} \, .  
\ee 
For $M_{BH} = 10^{20}$~g, this equation would then imply
that the total Schwinger luminosity is roughly
$10^{36}$~erg/s for a star of mass 10 -- 100~$M_\odot$, which
is comparable to the fusion luminosity for 10~$M_\odot$ stars given in
Eq.~(\ref{eq:lstar10}) but far below the fusion luminosity for 
100~$M_\odot$ stars given in Eq.~(\ref{eq:lstar}).
However this value of the Schwinger luminosity is
never reached, because the rate for proton capture is
$\sim 10^{29}$~s$^{-1}$ ($6 \times 10^{30}$~s$^{-1}$) for a 
100~$M_\odot$ (10~$M_\odot$) star, while the rate to create the 
$e^+/e^-$ pairs is the product of the production rate per unit
volume, $\sim m_e^4$, and the volume of the region around the BH in 
which the electric field exceeds the critical value $E_c$. The latter 
is a spherical shell of thickness about $1/m_e$, so the volume turns 
out to be $r_g^2/m_e$, where $r_g = 2 G_N M_{BH}$ is the BH 
gravitational radius. The pair production rate is
$\sim m_e^3 r_g^2 \sim  5 \times 10^{26}$~s$^{-1}$ 
for a $10^{20}$~g BH (the Schwinger discharge is fastest for this BH mass).
Hence the equilibrium between the capture and the Schwinger mechanism
is reached for a Schwinger luminosity that is several orders of 
magnitude lower than given above for the stellar mass 
$M_* = (10-100)$~$M_\odot$.
Thus fusion always dominates over the Schwinger effect as a heat source.

\section{Formation of a larger black hole at the center of the star via Dynamical Friction}

\subsection{Main Sequence Star}

The most important phenomenon associated with the PBHs inside the
first stars is the formation of a larger BH at the center. It is
well known that gravitational interactions cause every heavy body 
moving into a gas of lighter particles to lose energy by dynamical 
friction ({Binney} \& {Tremaine} (1987)). Thus, PBHs inside a star 
are expected to sink to the center of the star, eventually forming 
one single large BH. 

We will use Chandrasekhar's dynamical friction formula to compute the 
timescale for
the PBHs to sink to the center of the star.  If we assume that the gas of 
light particles
has a Maxwellian velocity distribution with dispersion $\sigma$, then the 
deceleration of 
a BH moving at a velocity $v_{BH}$ with 
respect to the rest frame of the fluid is
\begin{equation}
\label{eq:dynfric}
\frac{d}{dt}\vec{v}_{BH} = - 4 \pi \, G_N^2 \, M_{BH} \,  
\rho_b  \, \ln \Lambda \, 
\frac{\vec{v}_{BH}}{v_{BH}^3} \, \left[ {\rm erf}(X) - 
{2 X \exp(-X^2)\over  \sqrt{\pi}} \right]
\end{equation}
where $X \equiv v_{BH}/(\sqrt{2}\sigma)$, erf is the error function,
$\rho_b$ is the density of the background particles and
$\ln\Lambda \approx \ln(M_*/M_{BH})$ is the Coulomb 
logarithm~\footnote{The actual definition of Coulomb logarithm
is (see {Binney} \& {Tremaine} (1987))
\begin{displaymath}
\ln\Lambda = \ln \frac{b_{max} \, \sigma^2}{G_N (M_{BH} + m)} \, , 
\end{displaymath}
where $b_{max}$ is the maximum impact parameter, $\sigma^2$ is the 
mean square velocity of the gas and $m$ the molecular weight.
Numerical simulations show that $b_{max}$ can be assumed of order
the orbital radius of the object, say $R$. Since 
$\sigma^2 \sim G_N M_*(R)/R$, a reasonable estimate of $\Lambda$ is 
$M_*(R)/M_{BH}$.}.
There are two possible regimes, depending on  whether $v_{BH}$ is larger
or smaller than the velocity dispersion inside the star,
$\sigma = \sqrt{T/m}$ ({Binney} \& {Tremaine} (1987)). 
Here $T$ is the local gas temperature and
$m$ is the molecular weight.  The factor in the square brackets
\be
F(v_{BH}) = 
\label{F-of-v}  {\rm erf}(X) - {2 X \exp (-X^2)\over  \sqrt{\pi}}
\ee
tends to unity for $v_{BH}\gg \sigma$ and tends to 
$v_{BH}^3/2\sqrt{2\pi}\sigma^3$ for $v_{BH}\ll \sigma$.

The vector equation~(\ref{eq:dynfric}) can be rewritten as the following 
two scalar equations:
\be
\ddot r &=& -\frac{M_* (r) G_N}{r^2} + \frac{J^2}{r^3} 
- \gamma (v_{BH}) \dot r,
\label{ddot-r}\\
\dot J &=& -  \gamma (v_{BH}) J,
\label{dot-J}
\ee
where $r$ is the distance of BH form the star center, 
$J = r^2 \dot \theta$ is the BH angular momentum per unit mass, 
$\theta$ is the azimuth angle, $v_{BH} = \sqrt{ \dot r^2 + J^2/r^2}$, 
\be
M_*(r) = \int_0^r d^3 r \rho_b(r) 
\label{M-of-r}
\ee
is the stellar mass inside radius $r$, and 
\be 
\gamma (v_{BH}) = 4\pi G_N^2 M_{BH} \rho_b \ln \Lambda \,
\frac{{\rm erf}(X) - {2 X \exp(-X^2) /\sqrt{\pi}}}{v_{BH}^3} .
\label{gamma-of-v}
\ee

Since the characteristic gravitational time  scale
\be
\tau_g = \sqrt{\frac{r^3}{M_* (r) G_N}} \sim 
\left(\frac{3}{4\pi\rho_b G_N }\right)^{1/2} \approx 1900 \, 
\left(\frac{1 \; {\rm g/cm}^3}{\rho_b} \right)^{1/2} \; {\rm s}
\label{tau-g}
\ee
is much shorter than the lower limit on the characteristic dynamical friction time scale
\be
\tau_{DF} = \frac{\sigma^3}{4\pi\,G_N^2 M_{BH} \rho_b \ln \Lambda }
\approx 5 \cdot 10^{10} \, 
\left(\frac{10^{24} {\rm g}}{M_{BH}}\right)
\left(\frac{\sigma}{3 \cdot 10^7 \; {\rm cm/s}}\right)^3
\left(\frac{1 \;\rm{g/cm}^3} {\rho_b}\right) 
\left(\frac{10}{\ln \Lambda}\right) \; {\rm s} \, ,
\label{tau-DF}
\ee
we can approximately solve eqs.~(\ref{ddot-r}, \ref{dot-J}) as 
follows\footnote{The reader might also be concerned whether we can neglect  the third term in eq.~(\ref{ddot-r}) when considering  the opposite limit as $v_{BH}$ goes to zero.  In  this case,  $\sigma$ goes to $v_{BH}$ in eq.~(\ref{tau-DF}).  Again,  the third term in eq.~(\ref{ddot-r}) is completely negligible
and even more so than when eq.~(\ref{tau-DF}) depended upon $\sigma$.}.
We may neglect the last term in the r.h.s. of eq.~(\ref{ddot-r}) and
assume approximate equality $J^2 \approx G_N M_*(r) r $. Using this 
result we can integrate eq.~(\ref{dot-J}), which now takes the form:
\be
\dot v_{BH} = - \frac{\sigma^3 F(v_{BH})}{v_{BH}^3 \tau_{DF}}
\, v_{BH} \,,
\label{dot-v}
\ee 
and calculate the time of capture of small BHs at the stellar center. 
The result depends upon the initial velocity of the BH which we may 
estimate assuming that the BH is on a circular orbit of radius $r$ 
determined by the stellar mass $M_*(r)$ interior to radius $r$, i.e.,   
$v_{BH} = \sqrt{G_N M_*(r) / r}$. We find that, in the outer regions 
of the star, $v_{BH} \gtrsim \sigma$, in which case eq.~(\ref{dot-v}) 
scales as $\dot{v}_{BH} = - \sigma^3/(\tau_{DF}v^2_{BH})$. In the 
inner regions of the star,  near the stellar center, we find the 
opposite limit of $v_{BH} \lesssim \sigma$, in which case 
eq.~({\ref{dot-v}) scales as 
$\dot{v}_{BH} = - v_{BH}/ (2\sqrt{2\pi} \tau_{DF})$ instead. Thus, 
in the latter case, the time of BH formation is about
\be\label{tau-f} 
\tau_f &\approx& 2\sqrt{2\pi} \tau_{DF} 
\ln \left( \frac{v_{BH}^{in}}{v_{BH}^{f}} \right)
\approx 2\sqrt{2\pi} \tau_{DF} 
\ln\left(\frac{R_{in}}{R_f}\right) \approx \nonumber\\
&\approx& 1.4 \cdot 10^4 \,
\left(\frac{10^{24} \; {\rm g}}{M_{BH}} \right)
\left(\frac{\sigma}{3 \cdot 10^7 \; {\rm cm/s}} \right)^3
\left(\frac{1 \; {\rm g/cm}^3}{\rho_b} \right)
\left(\frac{10}{\ln\Lambda} \right) \; {\rm yr} \, ,
\ee
where $v_{BH}^{in}$ is the initial PBH velocity, so 
$v_{BH}^{in} \approx \sigma$ and implies $R_{in} \sim 10^{10}$~cm, while
$v_{BH}^f$ is the final PBH velocity, when $R_f = 4 \cdot 10^2$~cm,
that is, when the orbit of the PBH is equal to the Schwarzschild radius
of the final BH.
In the case $v_{BH} \gtrsim \sigma$, the timescale becomes
\be
\tau_f \approx \frac{\tau_{DF}}{\sigma^3} 
\left(\frac{1}{v_{BH}^{f}} - \frac{1}{v_{BH}^{in}}\right) \, ,
\ee
which can be quite a bit longer than the one for the case 
$v_{BH} \lesssim \sigma$ for $v_{BH}^{in}$ moderately larger than 
$v_{BH}^{f}$. As shown later, this is not a problem, because 
we always have a sufficient number of PBHs at small radii, 
where $v_{BH} \lesssim \sigma$.

The case of very eccentric orbits does not significantly change  
the picture. A simple estimate can be obtained assuming radial motion
and constant matter density $\rho_b$. In the absence of dynamical friction,
the motion of the PBHs can be treated as an harmonic oscillator
with period $\tau_g$ and velocity $\sim (R_0/\tau_g) \cos (t/\tau_g)$,
where $R_0$ is the maximum distance from the center of the star and 
$t$ is the time. Since the maximum velocity exceeds $3 \cdot 10^7$~cm/s
for $R_0 > R_* \sim 10^{11}$~cm, the equation of motion of PBHs 
inside the radius $R_*$ is basically
\be
\ddot{r} \approx - \frac{\dot r}{\tau_{DF}} - \frac{r}{\tau_g^2}
\ee
The differential equation is that of an underdamped harmonic oscillator:
\be
r(t) \sim e^{- t/2 \tau_{DF}} \cos(\omega t + \delta) \, ,
\ee
where $\omega\approx 1/\tau_g $.   
We find $\tau_f=2 \tau_{DF}\ln\left(\frac{R_{in}}{R_f}\right) $, a timescale
 which is actually shorter than in the circular case.  
Thus we expect that the result in eq.~(\ref{tau-f}) is a reasonable estimate of the timescale.


To obtain more accurate
quantitative estimates of the dynamical friction timescale on which
the PBHs sink to the center of the star, we did numerical calculations assuming 
that the star
can be modeled as an $n=3$ polytrope, 
which is known to roughly reproduce the stellar 
properties of a star dominated by radiation pressure.  We can then obtain density and temperature profiles 
for a 100~$M_\odot$ star which are plotted in fig.~(\ref{fig-den-temp}).
If one does the full stellar structure of a star of a Pop. III star,  the exact answer is different than found assuming
a polytrope. The difference is on the order of at most tens of a percent.

Subsequently, we can now compute the timescale for the case of
$M_{BH} = 10^{24}$~g; luckily, The resultant timescale can easily be scaled to other BH 
masses since it is
inversely proportional to $M_{BH}$.  To be specific, we investigated the case of
 a 100~$M_\odot$ star.
We found that the transition from fast to slow BH velocity (relative to gas 
particle velocity)
takes place at
$R_c \sim 2 \cdot 10^{10}$~cm. As explained above, the dynamical friction for BH
outside of this radius
is proportional to $1/v^2_{BH}$, while, for smaller radii it is proportional 
to $v_{BH}$.  Roughly
50\% of the BHs are initially inside the radius $R_i = 1.4 \cdot 10^{11}$~cm; these BH 
take $1 \cdot 10^4$~yr or less to sink to $R_c$. (We have also computed the timescales for infall
for BH coming in from different initial radii $R_i$ to the same value of $R_c$; 
our results are shown in Table~(\ref{table}). Subsequently each
BH takes another $\sim 5 \cdot 10^4$~yr to sink from $R_c$ to 
$R_f = 4 \cdot 10^2$~cm. The latter is the Schwarzschild radius of the
final BH at the center of star. Thus the timescale for  half of the PBHs 
to form a single large BH at the center of the 100~$M_\odot$ star is roughly
\be\label{ts-cbh}
\tau_f = 6 \cdot 10^4 \,  
\left(\frac{10^{24} \; {\rm g}}{M_{BH}}\right) 
\; {\rm yr}  .
\ee
Thus for $M_{BH} > 10^{22}$~g, in a 100~$M_\odot$ star,
the timescale for the formation of a large central BH
is less than a million years, which can have a significant impact on the 
evolution of the star.
We note that, once the central BH mass is $\sim 10^{25}{\rm g}$, 
the (fastest possible) accretion timescale in Eq.~(\ref{timescale}) 
becomes comparable to the dynamical friction timescale Eq.~(\ref{ts-cbh}); 
the result of both effects is a single large BH inside the star.

If the mass of the star is 10~$M_\odot$, the sinking time is not significantly
different.

Additional PBHs from outside the star may also fall onto the central BH via dynamical friction.
For a baryon density profile that scales as $\rho_b(r) \propto r^{-2.3}$ outside the star,
we find that the dynamical friction timescale is
\begin{equation}\label{tau-df}
\tau_{DF} = 2 \times 10^{16} {\rm yr} \left({{\rm ln}\Lambda \over 10}\right) \left( {r_i \over 1 {\rm pc}} \right)^{1.85}
\left( {M_{BH} \over 10^{24} {\rm g} }\right ) ^{-1} ,
\end{equation}
where $r_i$ is the initial radius of the infalling PBH
and this equation has been computed in the fast BH regime with $\dot v_{BH} \propto v^{-2}_{BH}$.
Thus it takes a very long time for dynamical friction to be effective at pulling in BH from typical
radii in the minihalo.  From closer in, the timescale can be significantly shorter, e.g.,
it takes 150,000 years for a $10^{24}$~g PBH to go from $3 \times 10^{12}$~cm ($\sim 10$ times
the radius of the star) to the center.
However, the amount of mass in PBHs inside this radius is $2.4 \times 10^{29}$~g, more than an order
of magnitude less than the amount already in the star from Eq.~(\ref{eq:mtot}), and is therefore negligible.  Thus dynamical friction does not pull
in significant mass in PBHs from outside the star.

\begin{table}
\begin{center}
\begin{tabular}{|c|c|c|r|}
\hline
$R_i$ (cm) & $M_*(R_i)/M_*(R_*)$ & $\rho(r_i)$ (g/cm$^3$) & Time (yrs) \\
\hline
\hline
$4.7 \cdot 10^{11}$ & $1.0$     &  $0.5$	&  205,000 \\
\hline
$2.4 \cdot 10^{11}$ & $0.9$     &  $3.1$	&  30,000 \\
\hline
$1.4 \cdot 10^{11}$ & $0.5$     &  $8.7$	&  11,000 \\
\hline
$6.5 \cdot 10^{10}$ & $0.1$     &  $17.3$	&  5,000 \\
\hline
\end{tabular}
\end{center}
\caption{Numerical results of the timescale for BHs to move from a variety of initial radii
 $R_i$ to a smaller radius $R_c \sim 2 \cdot 10^{10}$~cm. 
$M_*(R_i)/M_*(R_*)$ is the fraction of the mass of the star
inside the radius $R_i$, which is equal to the initial fraction of PBHs
inside the radius $R_i$. $\rho(R_i)$ is the stellar density at $R_i$. The mass of the BH has been fixed to $10^{24} {\rm g}$}.
\label{table}
\end{table}

\begin{figure}[t]
\par
\begin{center}
\includegraphics[width=14cm,angle=0]{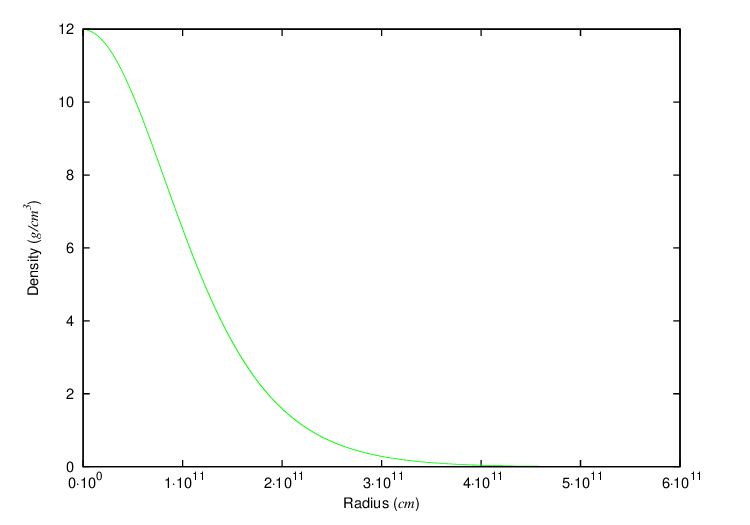}
\includegraphics[width=14cm,angle=0]{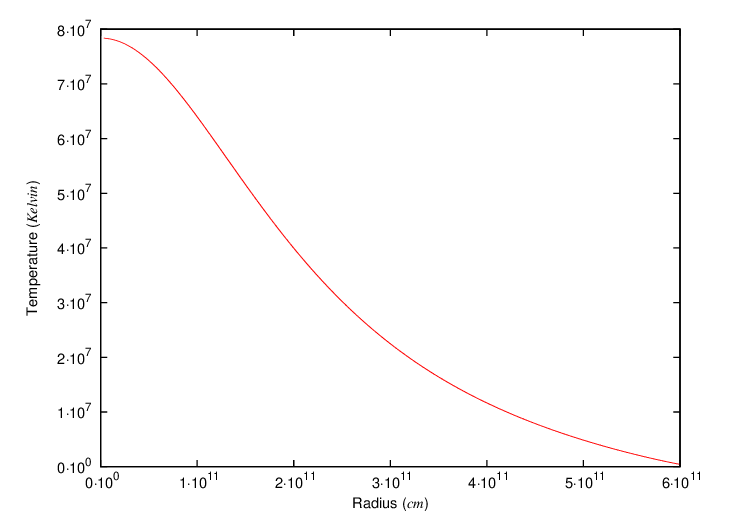}
\end{center}
\par
\vspace{-5mm} 
\caption{Density (top panel) and temperature (bottom panel) profile for the $n=3$ polytrope star of mass 100~$M_\odot$ used in our simulations.}
\label{fig-den-temp}
\end{figure}

\subsection{Protostellar Phase}

One may ask whether dynamical friction is already effective during the protostellar phase,
long before the Pop. III star comes to exist on the main sequence.  Early on, there is a collapsing
molecular cloud which is very diffuse and becomes more and more dense as it cools via molecular hydrogen cooling.  The protostellar
clouds stop collapsing once they become protostar nuggets with 
$10^{-2} \; M_\odot$ in mass, 
hydrogen densities of $10^{21} \; {\rm cm}^{-3}$, and radii 
$\sim 5 \times 10^{11}$~cm ({Yoshida} {et~al.} (2008)). 
In the standard
picture of Pop. III star formation, there is then accretion onto these nuggets until the stars reach
$\sim 100 \; M_\odot$ and go onto the main sequence.

Can the PBHs already sink to the center during this earlier phase and cause the protostar
to go directly to a BH, avoiding the main sequence phase altogether?  Inside the protostar,
the appropriate regime for dynamical friction is that of slow BH, with $\dot v_{BH} \propto v_{BH}$.
Such protostellar clouds have much lower densities than the subsequent Pop. III stars on the 
main sequence, and consequently are ineffective at causing the PBHs to slow down via
dynamical friction. We have checked that the timescale is simply too long for PBHs to play
any role during the collapse of the protostellar cloud, unless the PBHs are much more
massive than have been considered in this paper.  However, once the nugget forms,
the baryon density is high enough to trap PBHs of mass $> 10^{26}$~g with dynamical friction.  At this point the 
Kelvin-Helmholtz time $\sim \tau_{DF} \sim 10$~yr.   The amount of DM (PBHs) inside the nugget
is $\sim 10^{28}$~g, so that the initial central BH is only this big. However, 
it quickly eats the rest of the $10^{-2} \; M_\odot$ protostar, and presumably can grow at least to the value
of the original $1000\; M_\odot$ Jeans mass of unstable material.

\section{Eating Pop. III stars}

We have shown that the PBHs can sink to the center of the star and form
a single larger BH in a reasonable timescale (for $M_{PBH} > 10^{22}$~g)
to change the evolution of the star.
We now need to address the subsequent fate of the star: can the BH 
really accrete at the Bondi rate  and swallow the whole star quickly?  Alternatively,
does the radiation pressure from the accretion luminosity slow down the accretion rate and make the 
star have a normal evolution? In general, the accretion of matter onto
an object with a solid surface (e.g. a neutron star) is limited by the 
radiation produced by the accreting gas, 
\begin{equation}
\label{eq:eta}
L_a = \eta \dot{M}
\end{equation}
 where
$\eta$ is basically the gravitational potential per unit mass on the
surface of the object. Nevertheless, in the case of accretion onto
BH, the picture is more complex and the phenomenology richer. If the 
cooling mechanism is efficient, the gravitational energy of the accreting 
gas is radiated away and the gas temperature is much smaller than the 
local virial temperature. This case is similar to the one involving
objects with a solid surface: $\eta$ is equal to the binding energy
per unit mass at the Innermost Stable Circular Orbit (ISCO), since
we presume that the gas inside the ISCO falls quickly into the
BH and is unable to emit further radiation. So, for Schwarzschild 
BHs $\eta = 0.057$, while for Kerr BHs the efficiency parameter can be
as high as 0.42 ({Shapiro} \& {Teukolsky} (1983)). 
On the other hand, if the cooling is not 
efficient, the gravitational energy is stored as thermal energy into 
the gas rather than being radiated. That can occur if the gas density 
is very low and particles do not scatter each other very much, or in 
the opposite case, when the medium is optically thick and radiation is 
trapped, as happens for high accretion rate. Here, unlike neutron stars, 
BHs have an event horizon and the energy can be lost into the BH. $\eta$ 
turns out to be very small and the accretion luminosity can be low.
The accretion rate of matter can thus be high.

We will argue that the BHs at the center of the first stars may accrete at the Bondi rate, with
 the star adjusting to keep the luminosity equal to the Eddington value,
corresponding to a small value of $\eta$ in Eq.~(\ref{eq:eta}).
With Bondi accretion, the BHs can swallow the star in a short time, 
becoming 10--1000~$M_\odot$ BHs.
There is considerable discussion of BHs accreting material inside stars in the literature.
We present here some of the possibilities for the evolution of these objects.  In all cases,
the end result is a 10--1000~$M_\odot$ BH.
In the case of radiative stars, we may follow {Begelman} (1978), where
the author discusses the steady flow accretion onto a Schwarzschild 
BH of a non--relativistic gas where
the radiation pressure at infinity is much larger than the particle
pressure and the radiation--particle coupling is provided by the Thomson 
scattering. The medium is optically thick at all the relevant scales
and radiation is transported by diffusion and convection. Here one 
finds a trapping surface at the radius
\be
R_t = \frac{\dot{M}_{BH} \, \sigma_{Th}}{4 \pi m_p} \, ,
\ee
inside which the radiation is convected inward and swallowed by
the BH faster than it can escape to infinity.\footnote{Clearly $R_t$ cannot be larger than the radius of the star, $R_\star$.   In this case, we take $R_t= R_\star$.}
 If $R_t$ is much larger than the Bondi radiu $R_B=2GM/v^2$, then the radiation is effectively trapped, that is 
it is convected inwards faster than it can diffuse outwards. 
In our case, using eq(\ref{bondi-eq}),
\be
\frac{R_t}{R_B}=6\cdot 10^4 \left(\frac{M_{BH}}{10^{30} \; {\rm g}}\right)\left(\frac{\rho_b}{1 \; {\rm g/cm^3}}\right) 
\left(\frac{v}{3 \cdot 10^7 \; {\rm cm/s}}\right)^{-1} \, .
\ee

Given the typical densities and temperatures inside Pop. III stars, this condition is verified, 
the process is essentially adiabatic and in principle the BH is 
capable of accreting at an arbitrary high rate. Since radiation is trapped, 
the luminosity produced by the accretion process can not exceed the Eddington value, 
and the radiative efficiency effectively adjusts in order to keep 
$L\sim L_{Edd}$ ({Begelman} (1978)). 
As long as accretion is spherical, with zero angular momentum, the 
central PBH can accrete ad libitum, and eventually swallow the whole star.
In the presence of limited angular momentum we can argue that as long
as the accretion disk that forms is all contained within the trapping radius, 
then radiation remains trapped and the growth of the BH can continue 
({Volonteri} \& {Rees} (2005)).
We can take as a safe limit the condition that the disc is all within the trapping radius;
this provides a lower limit to when accretion stops.  The outer edge of
the accretion disk, $R_D$, is roughly where the specific angular momentum of the gas equals
the angular momentum of a gas element in a Keplerian circular orbit, therefore:
\be
\frac{R_D}{R_B}=\sqrt{2}\left(\frac{V(R_B)}{c_s}\right)^2
\ee 
where $c_s$ is the sound speed and $V(R_B)$ is the rotational component of the 
velocity at the Bondi radius. In this case it still seems  possible that the radiative efficiency 
drops so that the BH can accept the material without greatly exceeding the Eddington luminosity.
Relaxing the assumptions of zero angular momentum and absence of mechanical
turbulence and/or magnetic fields, the actual matter accretion
rate presumably decreases, but the evolution of the star is slowed
down as well. 
On the other hand, for very high angular momenta,
it sounds reasonable that the system looks like a collapsar 
({MacFadyen} \& {Woosley} (1999)).

{\it Convective stars:}
100~$M_\odot$ Pop.  III stars are primarily convective 
({Heger} {et~al.} (2007)). One can  compute the Eddington
luminosity in the case of a BH inside a mostly convective star with a
radiative outer envelope as follows ({Begelman} {et~al.} (2008)). 
There is no radiation pressure inside the
convective zone, so the luminosity from the BH can easily get to the
radiative outer envelope. Out there radiation pressure does
exist. Then the Eddington luminosity at this outer region
(which basically contains the entire star) is the relevant quantity. In short, one
should use the Eddington luminosity of the star rather than Eddington
luminosity of the BH, which means substituting $M_*$ for $M_{BH}$ in 
Eq.~(\ref{edd-lim}).
Doing this, one finds 
\begin{equation}
\label{eq:eddstar}
L_{BH} = L_{E,*} = 10^{40} \; {\rm erg/sec} \; (M_*/100 M_\odot) \, . 
\end{equation}
This value is significantly larger than the numbers obtained in 
Eq.~(\ref{edd-lim})
because it is the Eddington luminosity of the star rather than that of the BH.
Here the accretion
luminosity is bigger than the fusion luminosity. The consequence for
the star will be that it must expand, it will cool, and fusion will
shut off.  At that point the star looks like the quasistars in
{Begelman} {et~al.} (2008). These authors have worked out the stellar structure
for a BH of arbitrary mass inside a star of arbitrary mass, where the
only heat source is accretion luminosity. 
These authors were studying a different problem:  they were not looking at  Pop. III
 stars in $10^6 \; M_\odot$ haloes; instead they were looking at
 cooler regions of similar content in $10^7 \; M_\odot$  haloes. Although the 
 context was different, the resultant objects should be
 very similar.

 There are 2 possibilities for the
accretion: 1) The accretion may be spherical.  In that case $\eta$ can be
very small, as discussed in {Bisnovatyi-Kogan} \& {Lovelace} (2002).  
There is no problem having $\eta = 10^{-6}$ so that the Eddington
luminosity in Eq.~({\ref{eq:eddstar}) is compatible with  Bondi accretion at $\dot M = 10^{46}$~erg/sec.  Then it takes a thousand years to swallow
the 100~$M_\odot$ star (see Eq.~(53) in {Begelman} {et~al.} (2008)).  
Even more interesting is to contemplate the possibility that the star is
accreting further mass from the halo outside it, e.g. at a rate 
 $0.01\; M_\odot$/~year ({McKee} \& {Tan} (2007))\footnote{The accretion rate for Pop. III stars is still highly uncertain, and certainly variable as a function of time. The values we quote are higher than
  typical estimates for prolonged accretion rate (see e.g. Figure 8 of {McKee} \& {Tan} (2008)), but still definitely possible, especially if PBHs somewhat reduce feedback effects.}.  Then the BH can
end up very large as seen in Eq.~(52) of {Begelman} {et~al.} (2008), 
possibly eating all $10^5\; M_\odot$ of baryons in the DM halo.

 2) The accretion may be in a disk. If the disk is thin and radiatively efficient, then
 $\eta \sim 0.1$ and  $\dot M \ll \dot M_{B}$ (the accretion
 rate is much slower than Bondi).  
 However, in different geometries, $\eta$ can  
become much smaller ({Abramowicz} \& {Lasota} (1980)).
 {Begelman} {et~al.} (2008) claim that the accretion stops
 once you hit "the opacity crisis." This happens when the photospheric
 temperature (at the edge of the star) goes down to a critical value,
 so that the radiation pressure in the outer envelope vanishes,
 nothing prevents the star from going super--Eddington and blowing off
 all its gas.  This leaves behind an exposed BH that no longer
 accretes anything.
 They find that for a fixed stellar mass of 100~$M_\odot$, the resultant object is
  a 10 solar mass BH in $10^7$~years, but nothing bigger, due to this
 opacity crisis.  On the other hand, if the star is accreting further
 material from the outside, then you can end up with a 400~$M_\odot$
 BH if the accretion rate of material onto the star is
 $10^{-2}\;M_\odot/$~yr ({McKee} \& {Tan} (2007)) 
or 4000~$M_\odot$ BH if the accretion rate onto
 the star is $10^{-1}\; M_\odot/$~yr.  Again, it takes $10^7$~years to
 reach this.  In the meantime, during this $10^7$~years, you have a "PBH
 Dark (matter powered) Star", i.e.  a Pop. III star powered by accretion luminosity
 rather than by fusion. 
The  exact accretion rate is none the less quite uncertain. 
 Convective energy transport 
is itself limited and bounds the accretion rate
({Begelman} {et~al.} (2008))
\be
\dot{M}_{BH} \lesssim \frac{\dot{M}_B c_s^2}{\eta} \, .  
\ee
Since $c_s \sim 10^{-3} - 10^{-2}$, the actual accretion rate might 
be much smaller than the Bondi rate $\dot{M}_B$, unless $\eta$ is 
quite small, say $\eta < 10^{-4} - 10^{-6}$. This is not a problem
for spherical accretion, but might affect results for disk accretion.  Regardless, this will require more study.  Even accretion onto the first stars without the additional effects 
from primordial black holes  is
presently still an unsolved problem.

We have argued that the BHs at the center of the first stars may accrete at the Bondi rate,
so that the BHs can swallow the star in a short time, 
becoming 10--1000~$M_\odot$
BHs.  This mechanism may produce the seeds to generate the super--massive 
BHs  which have been observed even at high 
redshifts and at the centers of galaxies. 

\section{Summary and conclusions}

Primordial black holes in the mass range 
$M_{PBH} \sim 10^{17} - 10^{26}$~g are viable dark matter candidates. 
They may be produced in the early Universe by many mechanisms and so 
far there are no constraints on their possible abundance. Assuming 
that they make part of the cosmological dark matter, we expect that 
due to dynamical friction primordial black holes will make up a small 
but significant mass fraction of the first stars.  Primordial 
black holes with masses smaller than about $10^{22}$~g do not have
a significant effect on the evolution of primordial stars, because
their timescales for Bondi accretion and for dynamical friction are
larger than the lifetime of a Main Sequence star of $10 - 100 \; M_\odot$.
On the contrary, primordial black holes heavier than $10^{22}$~g
might sink quickly to the center of the star by dynamical friction 
and form a larger black hole, which could swallow the whole star 
in a short 
time. So, Pop. III stars would likely 
have lived for a short time, with 
implications for the reionization of the Universe after the cosmic 
dark ages and  the nature of the first supernovae; in fact they may preclude any 
supernovae from the first stars.  Although the BH swallowing the 
star shortens the star's lifetime and its contribution to reionization, 
the newly formed hole can become a new, alternative source of ionizing 
photons. The 10--100~$M_\odot$ BHs that form by swallowing the Pop. III 
stars may grow even larger: they reside in 1000~$M_\odot$ of gas that 
are in excess of the Jeans mass and may fall into the BH.  Black holes
of mass 1-1000 $M_\odot$ may result.

Depending on the accretion mechanism at this point,
the black hole may accrete  more matter and grow larger.    
The $10^6 \; M_\odot$ minihaloes of dark matter contain 
$\sim 10^5 \; M_\odot$ of baryonic matter.
This accretion from the minihalo, as well as from other haloes merging with the
one containing the black hole,  would be from low density gaseous material
($\rho \sim 10^{-24}$~g/cm$^3$),
which is considerably different from the accretion we considered earlier from  within
the star ($\rho \sim 1$~g/cm$^3$).  In the case of accretion from the low density gas
outside the star,
feedback may become important.  As we have shown, the 
timescale for the Pop. III stars to 
become black hole can be much shorter than the lifetime of the Pop. III stars 
(3~Myr for a 100~$M_\odot$ star), so that the feedback due to stellar 
heating and ionization of the medium surrounding
the black hole may be minimal. However, the
accretion may well be in a disk, with the accompanying radiation
pressure as well as radiative feedback due to the 
accretion ({Silk} \& {Rees} (1998); {Springel} {et~al.} (2005); {Ciotto} \& {Ostriker} (2001); {Li} {et~al.} (2007); {Pelupessy} {et~al.} (2007); {Alvarez} {et~al.} (2008))
limiting the accretion rate.    A recent study ({Alvarez} {et~al.} (2008)) 
of the radiative feedback from the black hole accretion
has been done for the case of $\eta = 0.1$ and a Pop. III star that has undergone
its full lifespan, and finds reduced accretion onto the BH; 
the story may be different here.   We have not studied these
later stages.  
Since the end--products are $10-10^5\; M_\odot$ black holes, these
objects may serve as seeds for Intermediate Mass Black Holes; the super--massive black holes
which have been seen already at high redshifts ({Haiman} \& {Loeb} (2001); 
{Volonteri} \& {Rees} (2006)) and may be
the progenitors of the super--massive black holes which are  in
the center of every normal galaxy today.

Even if the primordial black holes do not 
explain the entirety of the
dark matter in the Universe, they may still play a role in the first stars.
Heavier primordial black holes than the ones studied here, i.e., 
primordial black holes with $M_{BH} >  10^{26}$~g,  are 
observationally constrained to be only a fraction of the total 
dark matter in the Universe, and yet could be important in the first stars. 
 It would only take one such black hole to be pulled into the star via 
dynamical friction (timescale $\sim 10^7$~yr for a 1~$M_\odot$ 
black hole to get from 1~pc out into the center of the star (see 
Eq.~(\ref{tau-df})) and to quickly eat up 
the whole star. In fact, a single massive primordial black hole would
already have a major effect during the protostellar phase  while the 
molecular cloud is collapsing down into a protostar: the molecular cloud would
already collapse into a black hole. In this case the 
fusion phase of a Pop. III star would be completely avoided.   
Another possibility would be to have the dark matter consist 
primarily of Weakly Interacting Massive Particles (WIMPs) but with a 
small component of primordial black holes. In that case there would be 
dark stars powered by WIMP annihilation ({Spolyar} {et~al.} (2008)), 
which would become black holes once the 
primordial black holes sink to the center 
of the dark star.  

In principle, if the
effects described in this paper do not take place, one could place
bounds on the black hole abundances of various masses. 
For example, if primordial black holes swallowed 
primordial stars too quickly, the cosmological metal enrichment would be 
problematic and in absence of viable alternatives, the current 
allowed mass range $M_{PBH} \sim 10^{17} - 10^{26}$~g could 
be further reduced to $\sim 10^{17} - 10^{22}$~g.

\section*{Acknowledgments}
C.B., A.D. and D.S. thank  the Michigan Center for Theoretical Physics 
(MCTP) for hospitality during their visits in July 2008, when this work 
was started. 
C.B. was supported by the World Premier International Research Center 
Initiative (WPI Initiative), MEXT, Japan. 
K.F. was supported in
part by the DOE under grant DOE-FG02-95ER40899.
D.S. was supported by a GAANN fellowship.

\section*{References}

{Abel} T., {Bryan} G.~L. \& {Norman} M.~L., 2002, Science, 295, 93 \\
{Abramowicz} M.~A. et al. 2008, arXiv:0810.3140 \\
{Abramowicz} M.~A. \& {Lasota} J.~P., 1980, Acta Astron., 30, 35 \\
{Alcock} C. et al. 2000, ApJ, 542, 281 \\
{Alvarez} M.~A., {Wise} J.~H. \& {Abel} T., 2008, arXiv:0811.0820 \\
{Bambi} C., {Dolgov} A.~D. \& {Petrov} A.~A., 2008a, PLB, 670, 174 \\
---, 2008b, arXiv:0806.3440 \\
{Barkana} R. \& {Loeb} A., 2001, Phys. Rep., 349, 125 \\
{Begelman} M.~C., 1978, MNRAS, 184, 53 \\
{Begelman} M.~C., {Rossi} E.~M. \& {Armitage} P.~J., 2008, MNRAS, 387, 1649 \\
{Binney} J. \& {Tremaine} S., 1987, {\it Galactic Dynamics}
(Princeton University Press, Princeton, USA) \\
{Bisnovatyi-Kogan} G.~S. \& {Lovelace} R.~V.~E., 2002, 
New Astron.Rev., 45, 663 \\
{Bondi} H., 1952, MNRAS, 112, 195 \\
{Bromm} V., Coppi P.~S. \& {Larson} R.~B., 2002, ApJ, 564, 23 \\
{Bromm} V. \& {Larson} R.~B., 2004, ARAA, 42, 79 \\
{Carr} B.~J., 2003, Lect. Notes Phys., 631, 301 \\
{Carr} B.~J. \& {Hawking} S.~W., 1974, MNRAS, 168, 399 \\
{Carr} B.~J. \& {Sakellariadou} M., 1999, ApJ, 516, 195 \\
{Ciotti} L. \& {Ostriker} J.~P., 2001, ApJ, 551, 131 \\
{Crawford} M. \& {Schramm} D.~N., 1982, Nature, 298, 538 \\
{Dokuchaev} V.~I., {Eroshenko} Yu.~N. \& {Rubin} S.~G., 2007, 
arXiv:0709.0070 \\
{Dolgov} A.~D., {Kawasaki} M. \& {Kevlishvili} N., 2008, arXiv:0806.2986 \\
{Dolgov} A. \& {Silk} J., PRD, 1993, 47, 4244 \\
{Freese} K., {Spolyar} D. \& {Aguirre} A., 2008, JCAP 0811:014 \\
{Freese} K., {Gondolo} P., {Sellwood} J., \& {Spolyar} D. 2009, ApJ, .693, 1563 \\
{Gould} A., 1992, ApJ, 386, L5 \\
{Haiman} Z. \& {Loeb} A., 2001, ApJ, 552, 459 \\
{Hawke} I. \& {Stewart} J.~M., 2002, CQG, 19, 3687 \\
{Hawking} S.~W., 1971, MNRAS, 152, 75 \\
---, 1975, Comm. Math. Phys., 43, 199 \\
---, 1976, Comm. Math. Phys., 46, 206 \\
---, 1989, PLB, 231, 237 \\
{Heger} A. et al 2007, arXiv:0711.1195 \\
{Heger} A. \& {Woosley} S., private communication \\
{Hollenbach} D. \& {McKee} C.~F., 1979, ApJS, 41, 555 \\
{Jedamzik} K., PRD, 1997, 55, 5871 \\
{Li} Y. et al. 2007, ApJ, 655, 187 \\
{MacFadyen} A. \& {Woosley} S.~E., 1999, ApJ, 524, 262 \\
{Marani} G.~F. et al. 1999, ApJ, 512, L13 \\
{Matsuda} T., {Sato} H. \& {Takeda} H., 1971, Prog. Theor. Phys., 46, 416 \\
{McKee} C.~F. \& {Tan} J.~C., 2008, ApJ, 681, 771 \\
{Navarro} J.~F., {Frenk} C.~S. \& White S.~D.~M. 1997, ApJ, 490, 493 \\
{Nemiroff} R.~J. \& {Gould} A., 1995, ApJ, 452, L111 \\
{Novikov} I.~D. et al. 1979, A\&A, 80, 104 \\
{Omukai} K. \& {Nishi} R., 1998, ApJ, 508, 141 \\
{Page} D.~N., 1976, PRD, 13, 198 \\
{Page} D.~N. \& {Hawking} S.~W., 1976, ApJ, 206, 1 \\
{Peebles} P.~J.~E. \& {Dicke} R.~H., 1968, ApJ, 154, 891 \\
{Pelupessy} F.~I., {Di Matteo} T. \& {Ciardi} B., 2007, ApJ, 665, 107 \\
{Polnarev} A. \& {Zembowicz} R., 1991, PRD, 43, 1106 \\
{Ripamonti} E. \& {Abel} T., 2005, arXiv:astro-ph/0507130 \\
{Rubin} S.~G., {Khlopov} M.~Y. \& {Sakharov} A.~S., 
2000, Grav. Cosmol., S6, 51 \\
{Sellwood} J.~A. \& {MaGaugh} S.~S., 2005, ApJ, 634, 70 \\
{Seto} N. \& {Cooray} A., 2004, PRD, 70, 063512 \\
{Shapiro} S.~L. \& {Teukolsky} S.~A., 
1983, {\it Black Holes, White Dwarfs, and Neutron Stars}
(John Wiley, New York, USA) \\
{Silk} J. \& {Rees} M.~J, 1998, A\&A , 331, L1 \\
{Spolyar} D., {Freese} K. \& {Gondolo} P., 2008, PRL, 100, 051101 \\
{Springel} V., {Di Matteo} T. \& {Hernquist} L., 2005, MNRAS, 361, 776 \\
{Tegmark} M. et al. 1997, ApJ, 474, 1 \\
{Tisserand} P. et al. 2007, A\&A, 469, 387 \\
{Volonteri} M. \& {Rees} M.~J., 2005, ApJ, 633, 624 \\
---, 2006, ApJ, 650, 669 \\
{Yoshida} N., {Omukai} K. \& {Hernquist} L., 2008, Science, 321, 669 \\
{Yoshida} N., {Sugiyama} N. \& {Hernquist} L., 2003, MNRAS, 344, 481 \\
{Zeldovich} Ya.~B. \& {Novikov} I.~D., 1966, Sov. Astron., 10, 602

\end{document}